\documentclass[nonacm, sigconf]{acmart}
\usepackage{siunitx}
\AtBeginDocument{%
  \providecommand\BibTeX{{%
    \normalfont B\kern-0.5em{\scshape i\kern-0.25em b}\kern-0.8em\TeX}}}

\setcopyright{none}
\renewcommand\footnotetextcopyrightpermission[1]{} %
\usepackage{tabularray}

\begin{document}

\title{Make Making Sustainable: Exploring Sustainability Practices, Challenges, and Opportunities in Making Activities}

\author{Zeyu Yan}
\email{zeyuy@umd.edu}
\affiliation{%
  \institution{University of Maryland}
  \city{College Park}
  \state{Maryland}
  \country{USA}
}

\author{Mrunal Dhaygude}
\email{mrunald@umd.edu }
\affiliation{%
  \institution{University of Maryland}
  \city{College Park}
  \state{Maryland}
  \country{USA}
}

\author{Huaishu Peng}
\email{huaishu@umd.edu}
\affiliation{%
  \institution{University of Maryland}
  \city{College Park}
  \state{Maryland}
  \country{USA}
}

\begin{abstract}
The recent democratization of personal fabrication has significantly advanced the maker movement and reshaped applied research in HCI and beyond. However, this growth has also raised increasing sustainability concerns, as material waste is an inevitable byproduct of making and rapid prototyping. In this work, we examine the sustainability landscape within the modern maker community, focusing on grassroots makerspaces and maker-oriented research labs through in-depth interviews with diverse stakeholders involved in making and managing making-related activities. Our findings highlight four key themes: the various types of ``waste'' generated through the making process, the strategies (or lack thereof) for managing this waste, the motivations driving (un)sustainable practices, and the challenges faced. We synthesize these insights into design considerations and takeaways for technical HCI researchers and the broader community, focusing on future tools, infrastructures, and educational approaches to foster sustainable making.

\end{abstract}



\maketitle

\section{Introduction}

Over the past decade, the traditional relationship between consumers and mass production has evolved with the rise of affordable, compact, and user-friendly desktop fabrication tools, such as consumer-grade 3D printers and laser cutters~\cite{mota2011rise, lipson2010factory}. This shift is exemplified by the emergence of makerspaces globally~\cite{papavlasopoulou2017empirical}, the rise of numerous online 3D printing repositories (e.g., Thingiverse~\cite{thingiverse}), and the rapid growth of the desktop fabrication equipment market~\cite{lipson2010factory}. The democratization of these tools has also reshaped the landscape of applied research. For example, personal fabrication~\cite{baudisch2017personal} has become an important research domain within HCI, with many research labs transitioning from computing-focused spaces to ``maker'' spaces equipped with personal fabrication tools~\cite{barrett2015review}.

In many ways, this transformation parallels the rise of personal computing, where affordable, reliable, and accessible tools enabled individual developers—or, in the case of personal fabrication, individual makers—to innovate and express creativity~\cite{lipson2013fabricated, althaf2021evolution}.
However, personal fabrication introduces a distinct challenge that was not present during the democratization of personal computing: the environmental impact of material waste produced by makers. 
Traditionally, in mass production, waste consists primarily of packaging and electronic waste (e-waste)~\cite{tansel2017electronic}. 
While managing these waste streams has long been a persistent socio-technical issue, end users are generally not held responsible for the creation or management of this waste, apart from simple tasks like sorting it into appropriate disposal channels~\cite{islam2021global, shevchenko2019understanding}. 

With personal fabrication, however, the responsibility of waste generation is shifting directly to individual makers. 
As making becomes more accessible, the amount of material waste generated at the grassroots level is increasing. 
One example is the rise in thermoplastic waste—such as PLA, ABS, and PETG\footnote{PLA (Polylactic Acid), ABS (Acrylonitrile Butadiene Styrene), and PETG (Polyethylene Terephthalate Glycol) are common types of thermoplastics used in 3D printing.}—produced by desktop Fused Deposition Modeling (FDM) 3D printers.
In the UK alone, an estimated \SI{379000}{\kilogram} of plastic waste is generated annually from FDM 3D printing, with most of it inadequately processed or disposed of~\cite{filamentive}. 
As the push for democratized innovation continues~\cite{smith2009democratic, tanenbaum2013democratizing}, so does the unintended consequence of democratized waste generation~\cite{peeters2019barrier}.

This growing concern has recently drawn the attention of the technical HCI research community, leading to emerging concepts like ``unmaking''~\cite{Unmaking, 10.1145/3491101.3503721} or ``sustainable making''~\cite{yan2023future}. 
These ideas build on Blevis’s foundational work on sustainable interaction design~\cite{SID} and the establishment of sustainable HCI as a field, while extending much of the discussion into the domain of physical prototyping. 
For example, Song and Paulos ~\cite{10.1145/3544548.3581110} examine the ``afterlife'' of physical prototypes, using the lens of critical making to reconsider the relationship between the materiality of creativity and its environmental impact. 
Similarly, Lu and Lopes~\cite{lu2024unmaking} conduct a qualitative study, interviewing seven individual makers and artists who focus on reusing e-waste materials in various making activities. 
Their work urges HCI researchers to rethink the role of end-users in the broader context of sustainable computing.
In addition, there is a growing body of technical research addressing sustainability in personal fabrication. 
This includes investigations into 3D printing with waste-derived materials, such as coffee grounds~\cite{10.1145/3563657.3595983} and eggshells~\cite{bell2024demonstrating, dadhich2016simple}, studies on wearable prototyping using water-dissolvable yarns~\cite{lazaro2024desktop, zhu2024ecothreads}, and the development of new PCB prototyping techniques enabling the reuse of surface-mount electronics~\cite{yan2024solderlesspcb} as well as PCB substrates~\cite{yan2025pcbrenewal}.

Building on this body of literature, our paper contributes to the discourse on sustainable making~\cite{yan2023future}. 
However, rather than adopting a speculative approach to addressing material waste or framing unmaking as a conceptual lens for reflecting on making practices, our work is grounded in the current realities of maker culture and makerspace infrastructure in the context of sustainability. 
We aim to explore how today's makers perceive the sustainability impacts of their own making activities, the practices—if any—they adopt to reduce material waste, and the motivations and challenges they encounter in implementing sustainable approaches. 

Our qualitative research involves semi-structured interviews with 17 makers based in the U.S. 
We adopt a broad definition of ``maker'' as anyone with sufficient experience in creating physical artifacts and potentially generating material waste. 
This definition allows us to engage participants with diverse backgrounds, including managers and members of grassroots and university-based makerspaces, ``home'' makers with personal makerspaces (e.g., in their garages), and researchers whose work centers on making.

From our qualitative data, we identified four key themes at the intersection of making and sustainability: a detailed account of the types and sources of waste generated in common maker practices, the range of strategies (or absence of them) to manage this waste, the motivations behind adopting sustainable practices, such as feelings of guilt, and the challenges makers face, including the lack of adequate tools for recycling, reusing, or unmaking physical artifacts, even when they aspire to do so. 
Based on these findings, we organize the discussion around potential technical directions for addressing these challenges, with the goal of inspiring further research and promoting sustainable making practices within the HCI community.

\section{Related Work}
Our research builds on prior work in Sustainable Human-Computer Interaction (SHCI), the unmaking framework, and qualitative research on making and the maker movement. Each of these areas provides critical insights into how sustainability intersects with the practices and spaces of making.

\subsection{Sustainable HCI}

The concept of Sustainable Interaction Design (SID), introduced by Blevis~\cite{SID}, laid a foundational framework for what has now developed into the broader field of SHCI. 
Over the past decade, this field has grown to explore sustainability from multiple angles~\cite{10.1145/3411764.3445069, 10.1145/1753326.1753625}. Blevis’s original work emphasizes the need to integrate sustainability into every design phase, urging designers to reflect critically on how their values, methods, and decisions impact sustainability. 
This approach extends to material concerns such as the use, reuse, and disposal of products, with the ultimate goal of promoting longer product lifecycles through practices like repair and sharing.

Following this foundation, the theme of repair has emerged as a significant thread in SHCI, offering both theoretical and practical insights. Jackson and Kang, for example, conducted ethnographic studies of artists working with discarded technologies to challenge common assumptions about creativity and functionality~\cite{jackson2014breakdown}. Similarly, Lu and Lopes investigated e-waste practitioners, advocating for a reevaluation of users’ roles within capitalist systems~\cite{lu2024unmaking}. 
Kim and Paulos developed a framework for creative reuse, encouraging the transformation of e-waste through practices such as remaking and remanufacturing~\cite{kim2011practices}. 
Beyond creative reuse, repair itself has been studied in various contexts, such as Orr's ethnography of Xerox technicians, which highlights the collaborative nature of repair work~\cite{orr2016talking}, and research by Houston et al., which examined repair practices across diverse global settings~\cite{houston2016values}.

In parallel, SHCI research has explored persuasive technologies designed to influence end-user behavior, especially in relation to energy consumption. 
These technologies often use sensing and data visualization to encourage sustainable actions~\cite{10.1007/978-3-540-77006-0_7, 10.1145/3375182, 10.1145/1056808.1056932, 10.1145/3294109.3295634}. 
Mobile technologies have been particularly effective in promoting self-reflection and providing eco-feedback~\cite{10.1145/2786567.2795398, 10.1145/1753326.1753629, 10.1145/1518701.1518861}. 
However, much of this research positions end-users primarily as consumers, focusing mainly on goals such as reducing energy and water use, minimizing carbon footprints, and lowering electricity consumption.

In contrast to these consumer-oriented perspectives, our research explores sustainable practices among makers, who are not only consumers but also active creators of physical designs using tangible materials. This shift in focus allows us to examine sustainability from a different perspective, investigating how makers engage with materials and navigate sustainability in their daily activities, often without explicitly prioritizing material reduction.

\subsection{Unmaking and Sustainable Making}

The concept of unmaking, an extension of SID, offers a critical lens for examining the rapid production cycles driven by capitalism. 
The prefix \textit{un} suggests a deliberate rethinking of the assumption that products are inherently designed for longevity, exposing the realities of planned obsolescence, disposal, and replacement under the guise of technical advancement. 
This theme was recently highlighted as a special topic in TOCHI~\cite{song2025unmaking} and has been explored in projects such as Unmaking~\cite{Unmaking}, Un-crafting~\cite{murer2015crafting}, and Unfabricate~\cite{10.1145/3313831.3376227}, which use speculative and participatory design methods to investigate the afterlives of objects and materials.

For example, Song and Paulos experimented with 3D printing to explore how objects could have dynamic afterlives through unmaking~\cite{Unmaking}, while Sabie et al. used participatory design to foster critical conversations around unmaking as a way to provoke dialogue and embrace diverse perspectives~\cite{sabie2022unmaking}. 
Khan et al. discussed the pragmatics of sustainable unmaking through folk strategies involved in the e-waste recycling industry~\cite{khan2023pragmatics}. Meanwhile, Wu and Devendorf, through their Unfabricate project, reflected on the time and labor involved in the design-disassembly of smart yarn, framing this process as part of a broader critique of capitalist material production~\cite{10.1145/3313831.3376227}.

The connection between unmaking and sustainability is clear, particularly in initiatives such as the Sustainable Making workshop at UIST~\cite{yan2023future} and the Sustainable Unmaking workshop at CHI~\cite{song2024sustainable}. 
Our work aligns with these efforts but shifts the focus away from theoretical critique. 
Instead, we provide an empirical account of how sustainable practices in making are voluntarily implemented—or why they are not. 
This first-hand perspective offers valuable insights for future research aimed at promoting sustainable making practices at large.

\subsection{Make, Maker, and Makerspace}

The democratization of personal fabrication tools has transformed the landscape of making, leading to the rapid expansion of makerspaces and revitalizing maker and DIY cultures. 
Initially, these spaces served as grassroots hubs for technologists and hobbyists to engage in hands-on fabrication and hacking~\cite{lindtner2014emerging}.
Over time, the concept of a makerspace has broadened to include a wide range of physical spaces dedicated to making, from libraries to classrooms~\cite{hira2014classroom}, and even spaces supported by governments, businesses, or universities~\cite{freeman2018bottom, matthiesen2015replacing}.

In the HCI literature, the rise of makerspaces and maker culture has been extensively documented~\cite{lindtner2014emerging, bardzell2014now, tanenbaum2013democratizing}. 
For example, Shewbridge et al. explored the early uses of 3D printers in home makerspaces~\cite{shewbridge2014everyday}, while Hudson et al.~\cite{hudson2016understanding} examined how casual makers engage with 3D printing services in public makerspaces. 
Additionally, HCI research has highlighted the supportive and collaborative nature of these spaces. 
Kolko et al. documented Hackademia, a semi-formal learning environment that fosters creative engineering experiences through making in a shared, welcoming setting~\cite{kolko2012hackademia}. 
Similar findings emerged from a 19-month-long ethnographic study conducted by Toombs et al.~\cite{toombs2015proper}.

The growth of maker culture has positioned makerspaces as ideal environments for researchers to investigate the material and environmental aspects of physical creation. 
For example, Dew and Rosner~\cite{10.1145/3322276.3322320} presented a series of design studies that aim to understand how designers conceptualize, manage, and repurpose waste materials within on-campus makerspaces. 
Vyas et al.~\cite{vyas2023democratizing} conducted maker workshops with under-resourced communities to explore how e-waste can be leveraged to support engagement in technology design. 
Research outside of HCI has also looked into the intertwining of making, sustainability, and materiality. 
For example, research in environmental science has analyzed data from two separate makerspace printing locations to assess the uncertainties and variations in energy and material balances associated with democratized 3D printing~\cite{song2019uncertainty}.

Perhaps the most pertinent study related to our work is by Unterfrauner et al., as discussed in~\cite{unterfrauner2017environmental}. Their research, based on interviews with 39 makers across Europe, documented numerous maker projects and initiatives that either address environmental issues directly or adopt environmentally friendly making practices. 
For example, the study found that it is common for makers to use reclaimed pallets, repurpose scrap from different industries for prototyping, or utilize available household items to develop eco-friendly solutions for home appliances.

Our research also adopts a qualitative approach, involving semi-structured interviews with U.S.-based makers in various roles. Unlike previous studies, however, our work highlights many of the challenges associated with implementing sustainable making practices within today's maker infrastructure and provides design insights and considerations to address them as future research directions.

\section{Methods}
To investigate the interplay between sustainable practices and makers, we employed a semi-structured interview approach involving 17 participants, supplemented by field notes from on-site observations. Our methodology centered on conducting in-depth discussions to gain insights into makers' experiences. Following the interviews, we conducted a thematic analysis to systematically examine and interpret the data, allowing us to identify key themes and behavioral patterns within the context of making.

\subsection{Participants}
We recruited 17 makers based in the U.S. 
Participants were required to have a minimum of two years of experience in self-identified hardware prototyping or making activities. 
To ensure a diverse participant pool, we intentionally adopted a broad definition of ``maker.'' 
As a result, our participants included makerspace managers, individual makers such as hobbyists or entrepreneurs, university researchers, and those who combined these roles.
This diversity of roles allowed us to explore a range of perspectives---from those who primarily focus on personal making projects to those who manage the day-to-day operations of a makerspace, such as purchasing and organizing materials.

Recruitment involved a combination of purposive and convenience sampling. 
Purposive sampling was primarily used to recruit researchers and university-based makerspace managers from the authors' professional networks. 
These individuals were contacted directly and informed about the study's purpose and requirements.
Concurrently, convenience sampling was used to recruit participants from public makerspaces, including managers and makerspace members.
The authors regularly visited five local makerspaces, distributing recruitment advertisements through makerspace mailing lists. Interested individuals were then contacted for further communication and recruitment.

To build rapport with makerspace managers and makers, we actively engaged with them during visits and participated in makerspace events prior to conducting interviews. For example, at each makerspace we visited, we scheduled hour-long meetings with makerspace organizers, board members, or founders to introduce ourselves and explain our intentions. 
Additionally, we attended regular meet-up events open to non-members and participated in STEM educational exhibitions hosted by partner makerspaces. 
This approach allowed us to become part of their community, enabling us to ask more focused and informed questions about their activities and the makerspace environment during interviews and observations. 
A detailed demographic breakdown of the participants is provided in Table~\ref{tab:table1}.

\begin{table*}[ht]\label{table1}
\caption{Background Information of All Participants}
\begin{tabular*}{\linewidth}{@{\extracolsep{\fill}}lllllll}
\hline
\textbf{ID}   & \textbf{Role} & \textbf{Workspace} & \textbf{Experience} & \textbf{Format} \\ 
\hline
P1                 & Maker \& Researcher         & Research Lab                  & 5-10 years  & Online      \\ 
P2                 & Maker                       & Public Makerspace             & 10+ years  & On-site \\ 
P3                 & Manager                     & University Makerspace         & 10+ years      & On-site\\ 
P4                 & Maker \& Researcher         & Research Lab                  & 2-5 years         & On-site\\ 
P5                 & Manager                     & Public Makerspace             & 10+ years    &On-site   \\ 
P6                 & Maker \& Researcher         & Research Lab                  & 2-5 years    &Online   \\ 
P7                 & Maker                       & Public Makerspace             & 5-10 years    &On-site     \\ 
P8                 & Maker \& Researcher         & Research Lab                  & 2-5 years   &On-site       \\ 
P9                 & Maker \& Researcher         & Research Lab                  & 2-5 years  &Online        \\ 
P10                & Manager                     & Personal Makerspace           & 10+ years     &On-site    \\ 
P11                & Manager                       & Public Makerspace             & 10+ years    &On-site      \\ 
P12                & Manager                     & Research Lab                  & 5-10 years    &   Online  \\ 
P13                & Manager                     & Personal Makerspace            & 5-10 years  & Online  \\ 
P14                & Maker                       & University Makerspace         & 2-5 years  &On-site      \\ 
P15                & Maker                       & Public Makerspace             & 2-5 years    &Online     \\
P16                & Maker                       & University Makerspace         & 2-5 years  &On-site       \\
P17                & Maker                       & Public Makerspace             & 5-10 years   &On-site      \\ \hline
\end{tabular*}

\raggedright \footnotesize \textbf{*Note:} In the rest of this paper, we use the combination of [ID, Role, Workspace], as specified in the first three columns of Table~\ref{tab:table1}, to refer to a participant when presenting interview responses quoted from them. 
For example: [P3 - Manager - University Makerspace].
\label{tab:table1}
\end{table*}

\subsection{Procedure}
The interviews were conducted in a semi-structured format. 
When participants were located outside the authors’ state, the interviews were held remotely via Zoom. 
Otherwise, they took place in person, either at local makerspaces or at participants’ personal makerspaces. 
For Zoom interviews, participants were asked to join the call from their making environment so they could refer to their tools or projects if needed. 
During in-person interviews, on-site observations were also carried out, with the authors taking observational notes. 
Each interview lasted approximately 60 to 120 minutes, and participants were compensated at a rate of \$20 per hour. 

The interview questions were tailored to the role of each participant to ensure relevance. 
For example, researchers and makerspace members were asked about their previous projects, design workflows, iteration techniques, material choices, sustainability approaches, and project outcomes. 
These interviews were deeply rooted in the participants' own experiences, providing a detailed understanding of their processes. 
For makerspace and research lab managers, we included additional questions related to the daily management of the space, including equipment procurement, maintenance, upgrades, and waste management practices throughout the makerspace.

Our observation notes included physical layout and environmental cues, such as the location of recycling bins, visible waste within the space, and signage or posters promoting sustainability. 
The interviews were recorded and transcribed by the first two authors, and the field notes were documented and later reviewed to complement and contextualize the interview data.

\subsection{Data Analysis}

In total, we transcribed 27 hours of interview recordings. 
These transcriptions and field notes were used for an inductive thematic analysis~\cite{clarke2017thematic}. 
Specifically, the first two authors conducted multiple rounds of open coding without predetermined codes, allowing them to emerge organically throughout the analysis. 
The authors met regularly to review emerging codes, develop a codebook, and iteratively refine it. 
Through careful review and collaboration, a final codebook was developed, which was organized into a hierarchical structure of themes. 
A consensus was reached on key findings, which will be discussed in the following sections.

\subsection{Positionality Statement}

We are members of a research lab focused on interactive technologies, with expertise in hardware prototyping, and we strongly identify with the maker community. 
Reflecting on our own work and practices as makers, we were motivated to promote and integrate sustainable practices within this community. 
The research was conducted by the first two authors. 
The first author is an experienced fabrication researcher with an educational background in mechanical engineering and computer science, while the second author specializes in qualitative research methods with a background in electrical engineering and HCI. 
These diverse areas of expertise offer complementary perspectives on research methodologies. 
Our hands-on involvement in making has given us a deep understanding of the processes involved, which has been crucial in shaping the interviews and analyzing the data, enabling us to extract meaningful insights from the research.

\section{Findings}
In this section, we present four key themes from our findings to provide a comprehensive understanding of how material waste is generated, managed, and processed throughout the stages of making and prototyping—before, during, and after. 
Additionally, we examine the perspectives of stakeholders, both  individual makers and makerspace managers with varying resources, in addressing these material waste challenges.

\subsection{Types of Waste: That Which Is Wasteful and That Which Remains Usable}\label{type_of_waste}

We identified two common types of waste generated as byproducts of making: raw material waste produced during various machining processes or handcrafting, and artifacts, such as intermediate prototypes and archived projects, that may still be functional but are no longer in use.

\subsubsection{Material waste produced via various machining and handcrafting processes} \label{wastetypes}
Processed raw material created through making and prototyping is commonly identified as a significant source of waste. 
The type of raw material involved can vary depending on the project and the making process. 
We observed two common groups of material waste that were frequently referred to by our participants: thermoplastic in FDM types of 3D printing and crafting waste generated through various, often subtractive, machining processes.

Desktop 3D printers were present at all the sites we visited and in the research labs we interviewed, often described as central to any project involving the creation of a physical prototype. 
When asked about material waste in making, participants frequently identified thermoplastics such as PLA, ABS, and PETG as major sources of raw material waste. 
The bottom-up printing process inherently requires support structures, which become waste once printing is complete. 
Additionally, 3D printing failures are common, often resulting in significant amounts of plastic waste.

Beyond 3D printing, subtractive manufacturing methods such as CNC machining and laser cutting were also popular at the sites we visited. 
Unlike 3D printing, these machines can work with a broader range of raw materials, such as wood, acrylic, and metal sheets. 
The subtractive machining process produces two main types of waste: offcuts that are too small to be reused, and, in the case of CNC machining, swarf—sawdust and metal chips generated during the process. 
P7 stated:

\begin{quote}
\emph{``So, the one thing that I think is most wasteful in the shop is when I do acrylic prototyping using lasers to cut the acrylic. There's no way to recycle it, so you try to use as much of it as you can. Eventually, the corner pieces get too small to be useful for anything, and they're just not viable.''} [P7 - Maker - Public Makerspace]
\end{quote}

A final category of by-product waste that we uncovered included chemical solutions used in either the preprocessing or postprocessing of certain materials. 
Examples include the alcohol used for cleaning 3D printed models made of resin and the liquid used in desktop waterjet cutters.

\subsubsection{Brand-new components, intermediate prototypes, and archived projects}\label{second-type-waste} 
While byproducts generated through machining are, unsurprisingly, mostly waste, participants noted that functional components and hardware prototypes can also be considered waste. This includes brand-new components purchased before the start of a project to test new approaches, which may become unnecessary if the project's direction changes; intermediate prototypes created during the making and design process to verify specific features, which are no longer needed after testing; and finished projects that are not maintained and thus become obsolete.

P10 showed us a few brand-new custom PCB boards. 
These boards were placed in a small cardboard box next to his working desk, where he kept parts that were not needed:
\begin{quote}
    \emph{``I designed these PCB boards for this robot project because I needed them to be small and breadboard is too big. These PCBs are made by another manufacturer, but all these companies have a minimum order for PCBs. So, I ordered 10 of them and used 2 to test the design. The remaining 8 are still good, but now I don't need them.''} [P10 - Manager - Personal Makerspace]
\end{quote}

Similarly, P9 reported how he often orders what he needs during prototyping, but does not revisit them afterward:
\begin{quote}
\emph{``I usually buy different parts during prototyping because I don't know which one works. I need to do my experiment. If they don't work out, you're like, `Okay, returning it is too much work, so I guess I'll just keep it here.' And then nobody uses it, right? It ends up being a waste of money and it takes up space. Eventually, you're like, Okay, let's either toss this or give it away.''} [P9 - Maker \& Researcher - Research Lab]
\end{quote}

Intermediate prototypes or previously fabricated physical artifacts can also become material waste. 
This is especially true for complex projects that require several design iterations. 
P16 showed us a fairly large 3D printed part that was lying on his desk.  

\begin{quote}
    \emph{``I was designing an enclosure for a screen to post a home signage, and I went through a few iterations. The part was printed a few days ago, overnight. Then next day I found two problems arose and I couldn't use the printed part anymore. I now have these pieces lying around, and I have no use for them.''
    }[P16 - Maker - University Makerspace]
\end{quote}

\subsection{Handling Waste: Trashing, Keeping, and Reusing}\label{handling_waste}
As material consumption—and consequently, waste—is a routine part of making, both makerspaces and makers themselves have developed different strategies to streamline the process of handling it.

\subsubsection{``They are all going to the trash bin.''}\label{waste_swarf}
Most machining byproducts and daily consumables are directly discarded, making up a large portion of the overall waste. At several sites, we observed large trash bins installed close to fabrication machines. Material scraps, such as corner pieces of acrylic or CNC swarf, are mixed together and thrown directly into these bins.

\begin{quote}
\emph{``In the CNC shop, we have this rule where everyone has to vacuum after they finish their job. But the suction isn't 100\% efficient, you know? There are always bits lying around. It's just basic cleaning stuff. But I think all that waste just ends up in a landfill because you can't really recycle it. Mixing different materials makes it hard to recycle.''} [P5 - Manager - Public Makerspace]
\end{quote}

The handling of thermoplastic waste varies. For example, at some places, 3D printed waste is treated the same as CNC scraps and other raw material waste, all disposed of in the same trash bin. However, at other sites, such as some university-funded makerspaces and research labs, we observed that 3D printed waste had its own dedicated trash bin.

\begin{quote}
    \emph{
    ``We've got trash bins here, so after I finish printing, I come over here to clean up. Any supporting parts go straight into the bin.'' 
    } [P16 - Maker - University Makerspace]
\end{quote}

The collected 3D printed waste then goes through the recycling waste stream, with contractors collecting it periodically.

However, it should be noted that recycling 3D printing waste is not straightforward. In reality, not all 3D printing materials are the same from a recycling standpoint. PLA is the most commonly used 3D printing plastic filament and is easier to recycle than materials like PETG, ABS or composite filaments infused with, for example, carbon fiber. Since these materials require different thermal and chemical treatments during the recycling process, they should not be mixed together. However, according to several makers, printed waste is treated the same during disposal: it is neither sorted by type during 3D printing nor handled separately after printing is complete. Instead, it is indiscriminately discarded in the same dedicated 3D printing trash bin at their sites.

One practice we learned from a makerspace manager, who is more concerned with the recycling of thermoplastics, is to address this issue by only supporting PLA-based 3D printing.

\begin{quote}
\emph{``You can recycle PETG filament, but most places won't accept it. I try not to be the person who throws everything into the recycling bin and leaves it for them to sort out. Very few places actually recycle PETG parts because they often can't identify the type of plastic. In the end, we found that sticking to PLA only helps with recycling.''} [P11 - Manager - Public Makerspace]
\end{quote}

The last group of waste that we identified as being thrown away consists of components or prototypes that are still functioning. It's worth noting that not all prototypes end up in the trash bin. As we will explain in Section~\ref{storage}, some makers choose to keep as many prototypes as possible until their storage reaches its capacity. However, we also found that in different cases, makers may directly throw away functional components or even entire prototypes. In one example, P6 discussed the early assembly he made for his research, which involved metal components such as bolts and nuts used to put two 3D printed shells together with wires and a few LEDs embedded. The entire assembly was thrown away in a trash bin without being disassembled:

\begin{quote}
    \emph{``I know it has small bolts and nuts, and also wires. Now that you ask, it's better to remove them from the plastic before trashing them. But most of the time, I don't think about it but just throw them all together.''} [P6 - Maker \& Researcher - Research Lab]
\end{quote}

\subsubsection{``I wish I could keep them all.''}\label{storage}
For finished projects and large intermediate prototypes, storing them on-site is common practice. Several participants reported keeping most of their project iterations and intermediate prototypes even after the project had concluded. For them, these iterations signify the hard work that went into the project, and they want to have a memory of it. P7 said:

\begin{quote}
\emph{``I keep them as long as I can bring myself to do so. I also number all my prototypes. Sometimes, it's nice to look at all of them.''} [P7 - Maker - Public Makerspace]
\end{quote}

P1 also expressed a similar sentiment. When asked how he handles the prototypes he developed, P1 said:
\begin{quote}
\emph{``I kept some prototypes for that very reason, to have them on my desk,  because they were nice. And this was all my blood, sweat, and tears in there.''} [P1 - Maker \& Researcher - Research Lab]
\end{quote}

Besides archiving finished projects for documentation purposes, participants also mentioned storing items for potential future use. They reported keeping various intermediate prototypes and broken machines, hoping to salvage hardware or spare parts as needed in the future. P5 said:

\begin{quote}
\emph{``It depends on whether we can scavenge parts from it. So for example, we have this color thermal printer that we bought off eBay. It's broken and doesn't work, but we keep it around because we think there might be some interesting components in it.''} [P5 - Manager - Public Makerspace]
\end{quote}

Our data indicates that participants typically stop storing items or discard stored projects only when they run out of space. 
When this happens, it triggers a deep cleaning process. 
Lab or makerspace members will quickly sift through the parts. 
Since the primary goal is to free up space, this cleaning process can be rushed, and parts that were saved for future use may eventually end up in the trash bin.

P7, who tends to keep all his prototypes, admitted that this can be very challenging. 
When talking about the prototypes that he currently has, he said:
\begin{quote}
    \emph{``I mentioned being on the 35th revision of this project, but I don't actually have 35 of them. There is no space for me to keep all of them so some are gone during cleaning.''} [P7 - Maker - Public Makerspace]
\end{quote}

\subsubsection{``The motor got three more lives.''}\label{recycle}

While a great amount of waste is discarded, we encountered three approaches that practitioners have employed to find new uses for it. 

One approach is to reuse still-functional components, especially if they have a relatively high value or are otherwise difficult to find. 
P10 shared a story with us about how he salvaged a fully functional motor from a broken machine and repurposed it for different projects:

\begin{quote}
\emph{``I had a window regulator in my car that went bad. You know, I had to buy a new one 'cause the pulley was all messed up. But the motor was awesome, so I saved it. Two years later, it ended up in a gantry [of a DIY robot]. And then, as the robot grew, it wasn't big enough for that anymore. So that same motor got repurposed into something to spin a rotary test tool we're using. So yeah, that motor that came out of a car actually had three more lives after the car.''} 
[P10 - Manager - Personal Makerspace]
\end{quote}

The value of reusing functional components can be even greater for certain makerspaces, particularly if it helps reduce operational costs. 
P5, a public makerspace manager, mentioned how the high school robotics team, who are members of the space, improvises on their robot each year to meet the requirements of new design challenges, largely by reusing functional components from previous projects:

\begin{quote}
\emph{``They don't keep all the robots. They dismantle the previous one and use the same motors from the old one to build a new one. I also dismantle things that I know no one will use. If somebody needs an Arduino, I'm not going to buy a new one; I'll give them one I found from an old project.''} [P5 - Manager - Public Makerspace]
\end{quote}

In addition to being reused locally, components—and sometimes entire pieces of machine hardware—may circulate within the community. 
P5 shared that their makerspace often receives donations of hardware, including malfunctioning 3D printers, CNC machines, and sensors. 
Some of these donations are repaired and retained for makerspace operations, while others are rehomed and returned to the community.

\begin{quote} \emph{``I arrange something called `Fix it and its yours' [workshop]. Anyone who fixes the machine first gets to keep it. This way the machines don’t go out of use. We find them a new home.''} [P5 - Manager - Public Makerspace] \end{quote}

While electronic components and hardware machines have direct value in reuse, reusing material waste generated through machining or crafting processes is less straightforward. 
In such cases, the waste may be repurposed into physical art decorations if a maker recognizes its artistic potential. 
For example, P11 discussed his plan to turn boxes of discarded beer caps and cans he had collected into a mosaic, inspired by their vibrant array of colors. 
Similarly, P12 talked about his wall decorations made from 3D printed waste collected at his site, a project he was very proud of:

\begin{quote}
\emph{``This waste is from my work, and I tried to convert it into some art pieces. So, for example, in my own home, I created a small transparent bag and filled it with failed 3D printed materials. Next to the bag, I think I labeled a packet and printed something on it, and the whole thing will be like a poster. It’s a 3D poster. And I hang it on my wall.''} [P12 - Manager - Research Lab]
\end{quote}

A final approach to reusing material waste is through local recycling. 
P6, who works in a well-equipped university lab, discussed the practice of their university for recycling PLA waste within university facilities. 
In his lab, they follow a strict rule allowing only PLA from a specific brand. 
After printing, all PLA scraps must be sorted by color and then sent to professional recycling equipment, where they are re-extruded into new filament. 
While this approach makes great use of material waste, the process itself is costly and difficult to maintain:

\begin{quote}
\emph{``It's really nice that we can recycle PLA, but it really took them a long time to get to a consistent process, even with the setup they already have. They also need to do regular maintenance, like taking out the nozzle, bringing it to the workshop, and hitting it with a blowtorch just to burn off all the PLA. Then they clean the nozzle and put it back in. This is something they have to do every couple of weeks. The whole recycling process, it's taken them about a year to fine-tune it and get to a point where they can make consistent filament from just PLA, nothing else.''} [P6 - Maker \& Researcher - Research Lab]
\end{quote}

\subsection{Challenges of Being Sustainable}\label{challenges}
All participants in our study are well aware of both the importance and the potential environmental impacts of making. Through our interviews, many expressed concerns about frequently producing material waste. 
\textit{``I feel guilty about all the wood I use and waste in the process, so I try to plant trees equivalent to my usage.''} 
However, we found that incorporating sustainable making practices into daily activities poses three challenges: the lack of appropriate processes and equipment to handle waste materials, the high costs associated with sustainable practices, and the lack of technical know-how.

\subsubsection{Lack of procedures and equipment}\label{lack_equipment}
Participants discussed their attempts or intentions to be more sustainable. However, they cited a lack of practical guidance and tools within the community as obstacles to these efforts. For example, during our visit to P3's site, he referred to an entire pile of plywood and said:

\begin{quote}
\emph{``We use large amounts of plywood in the woodshop. 
And those are frustrating for us because I don't have a good recycling solution. All these sawdusts generated through CNC or laser cutting. 
We've talked to different people on campus, we've asked amongst other university makerspaces what ideas they have.
We haven't really gotten very far with it.''} [P3 - Manager - University Makerspace]
\end{quote}

The same concerns were raised regarding the inability to recycle printed plastic. P3 showed us their efforts to turn 3D printed waste into injection molds for reuse. 
However, the entire process requires a specialized device to melt or dissolve the plastic and the preparation of a separate mold each time. The process is not streamlined, and due to the need for precise temperature control during the melting phase, they haven't achieved satisfactory results. 

P11 also expressed concern about 3D printed waste. When discussing the daily waste produced though 3D printing, P11 commented:

\begin{quote}
\emph{``That's a lot of plastic. I really don't want to throw them away if I can have a way of using them, but there's not much you can do with with scrap 3D printed parts. Unless, I mean, you've got to have, like I was saying, gotta have a grinder and an extruder to turn it back into usable filament. But these are not like a cheap printer that you can just buy.''}
[P11 - Manager - Public Makerspace]
\end{quote}

The grinder and the extruder that P11 referred to are specialized devices capable of breaking down 3D printed plastic parts into smaller pellets, which can then be melted and remade into new filament. 
As discussed earlier in Section~\ref{recycle}, very few well-equipped makerspaces may have access to such machines. However, compared to the overall number of makers or the prevalence of 3D printing services within these makerspaces, these devices are not accessible to the majority.

\subsubsection{High costs in sustainable actions}
The cost of sustainable actions is another practical reason why makers and researchers don't always give them priority when making decisions. 
Being more sustainable often requires additional steps, extra labor, or increased costs, all of which can be burdensome for many.

For example, properly disassembling an intermediate prototype is a reasonable approach to reclaim functional components, as we identified in Section~\ref{recycle}. 
However, in practice, this process can be demanding, as it may require pausing ongoing design iterations and dedicating time and effort to tasks like unscrewing parts or desoldering electronics from an old board. People may choose to recycle components only when they deem it worthwhile. As P8 said:

\begin{quote}
\emph{``I think we try to balance between the time we spend on, like recycling or any of that, versus the actual research outcome. So, as a research lab, our time is valuable. Being more sustainable is not one of our brand, so if throwing away a very cheap thing is gonna save you a lot of time on research, then just do it.''} [P8 - Maker \& Researcher - Research Lab]
\end{quote}

Makerspace managers resonated. When asked how they decide what to disassemble or reuse, P10 responded:
\begin{quote}
\emph{``Depending on their value. But the screws? Typically not. Because we can buy a box of 100 screws for like \$4.
It's just way easier to buy new ones than spend more time or hire someone to dismantle and reuse existing ones.''} [P10 - Manager - Personal Makerspace]
\end{quote}

Similarly, P7, who used to collect a large amount of metal chips from CNC milling, explained why he didn’t want to bring them to a dedicated recycling facility:

\begin{quote}
\emph{``I'm not gonna get any money for this from the scrap yard. It's literally not worth my gas to get there and get back. And I don't want to damage the planet. I'm doing in my opinion worse by just wasting gas.''} [P7 - Maker - Public Makerspace]
\end{quote}

\subsubsection{Knowledge gap}\label{gap}
One additional concern is the lack of sufficient knowledge and understanding on sustainability. The democratization of digital fabrication tools has made design and prototyping more accessible than ever. However, individuals new to making may lack an understanding of appropriate methods for material handling and waste management. For example, at one site we visited, we observed a member improperly disposing of isopropyl alcohol---used for washing resin-printed parts---by pouring it directly into the sink. The individual was unaware of the issue until we pointed it out. P5 indicated:
\begin{quote}
\emph{``Whenever someone handles things incorrectly, someone knowledgeable might try to point it out. But I think things definitely slip under the radar, and it's difficult to keep track of everything.''}
[P5 - Manager - Public Makerspace]
\end{quote}

P3, a full-time makerspace manager for several years, reported similar difficulties in passing on the knowledge of recycling:
\begin{quote}
\emph{``One of the overall more general problems at a makerspace is people see a recycling bin, and they think they can put anything plastic in the recycling bin, right? 
Even though our recycling bins are pretty well labeled for what's supposed to go in there. 
And I have to constantly remind them, if it doesn't have a recycling stream number on the bottom, you don't put it in there.''} 
[P3 - Manager - University Makerspace]
\end{quote}

In other cases, a lack of knowledge about handling materials or components may also discourage recycling efforts. For example, makers may hope to recycle electronic components but may not possess the required skills or knowledge to do so. P6, an expert in electronics engineering, observed:
\begin{quote}
\emph{``Sometimes if things are soldered, for instance, a motor was soldered, right? 
So in this case, if we want to reuse the motor our only option is to desolder it, but not everyone here knows how to desolder. So a lot of times people do just buy new things because they don't know how to desolder.''}
[P6 - Maker \& Researcher - Research Lab]
\end{quote}

\subsection{Achieving Sustainable Making Through Indirect Means}\label{indirect_means}
While we have discussed the types of waste, current handling practices, and the challenges of maintaining a sustainable making process, it is important to note that there are also existing common practices that contribute to more sustainable ways of making—even if sustainability is not the primary goal. We argue that these practices are readily adoptable and should be encouraged community-wide.

\subsubsection{Optimizing design iterations}\label{optimize_iteration}
Almost all the makers discussed how their projects require design iterations and prototyping. 
These processes naturally generate a number of intermediate prototypes, which, as explained in Section~\ref{second-type-waste}, serve as a natural source of waste. 
Thus, optimizing design iterations and using fewer prototypes, ideally without compromising design goals, will result in reduced material usage, fewer physical artifacts, and more sustainable actions. 

In our interviews, several participants discussed their practices for reducing trial-and-error. 
For example, P1 explained how he uses design software and simulation to minimize the number of iterations needed for creating multiple physical prototypes:

\begin{quote}
\emph{``We would have had more tries without the simulations. I can't really say like, how much of an increase that would have off the top of my head. But I am confident that it will be way more iterations that we actually had to do.''} 
[P1 - Maker \& Researcher - Research Lab] 
\end{quote}

Other participants shared their strategies for optimizing design iterations, such as printing parts of a model rather than the whole, to save testing time, which also led to material savings. 
For example, P12 reflected on a project that involved 3D printing intricate spring structures with prismatic joints at their center:

\begin{quote}
\emph{``For the prismatic joint, we need to test out the tolerance inside the joints. I just print out the joint itself. 
I don't print the rest and I just test out this, the joints. It saves time this way. Print the entire thing will take much longer and I only need to test part of it so this is more effective.''} 
[P12 - Manager - Research Lab]
\end{quote}

While saving materials and reducing component use may not be a primary concern for participants during the design and prototyping process, many makers, especially those with extensive design experience, naturally optimize design iterations, contributing to practical approaches to more sustainable production.

\subsubsection{Better organization through labeling}
Throughout the study, we observed that nearly all sites had developed their own maintenance rules and practices. Participants reported that labeling was their go-to method for organizing the space—whether for tools, projects, shelves, or consumables. Labeling helps people locate items, even if they weren’t the ones who originally placed them. 

\begin{quote}
\emph{``I think one thing that makes a huge difference on whether things are put back and reused or not, is how organized things are and how sensitive the components are. 
So if there is a definite place for something to go, like people are more likely to, you know, once they unplug it, put it back in the workplace.''}
[P2 - Maker - Public Makerspace]
\end{quote}

One of the makerspace managers mentioned having dedicated, labeled shelves along with an instruction sheet explaining the labels in further detail. 
P3 had a unique way of labeling things with both textual and pictorial representations:

\begin{quote}
\emph{``We made this tool wall early on, this is one of the first things we did when I first came here, you know, we had these toolboxes, these tools. And I thought, well, that's we'll just put all the tools in those and we'll label the drawers. That'll work. And the volunteer students at the time said, Well, you know, that's kind of frustrating for us, because we don't really, you know, we'll look at the label that says whatever reciprocating saw, and what's a reciprocating saw all these saws reciprocate, right? So instead, we decided that we would do these tool boards where, you know, there's a picture of the tool behind it.''} 
[P3 - Manager - University Makerspace]
\end{quote}

While labeling is considered a common practice among makers, it can fall short in a complex and dynamic making space where novel or unconventional materials are brought in from time to time. 
P8 appreciated the exhaustive labeling system at their site for aiding in space organization and reuse, but he also shared occasions when the system failed:

\begin{quote}
\emph{``One of the projects recently needed PVC pipes, and there's no designated place for PVC pipes, which means that the structure that was built was disassembled, and the PVC pipes are just like lying around everywhere in the lab, instead of going back to the designated place. And like, you know, like being reused for other projects, because like people don't know that there are actually PVC pipes that they can use is just in random places everywhere around the lab.''}
[P8 - Maker \& Researcher - Research Lab]
\end{quote}

When asking about how these labels are created or kept tracking of, most sites rely on their managers and often do it manually.

\section{Discussions}

Our findings are the first to provide a holistic examination of the intricate landscape of sustainable making practices within the context of current makerspaces. We began by categorizing the different types of waste generated as byproducts of various making activities. For processes centered on digital fabrication or handcrafting, byproducts typically consist of raw material waste, whereas for processes focused on assembly, waste may include intermediate projects or even finished ones that are not perceived as useful at the moment (Section~\ref{type_of_waste}). Depending on their potential value, we identified three major practices for handling these byproducts. Projects or devices with high economic value or personal attachment are often reused or, at the very least, shelved for potential future use. However, in many cases, they are eventually discarded, often indiscriminately, alongside raw material waste (Section~\ref{handling_waste}). 

We found that makers are highly aware of the environmental impact of their activities, as all participants acknowledged—and at times felt guilty about—the unavoidable material waste generated during the design and making process. However, in current practice, makers reported major challenges in adopting more sustainable practices for handling wastes: the lack of equipment, the high trade-offs, and the limited know-how, particularly among new makers (Section~\ref{challenges}). Finally, we identified two groups of practices that, while not explicitly aimed at sustainability, contribute to it implicitly (Section~\ref{indirect_means}).

So, what do we learn from these findings? And how might these findings inform HCI research aimed at promoting more sustainable making practices?

One insight that may come as a surprise is the fluidity of waste, which is shaped not only by its objective functionality (e.g., whether it is malfunctioning) but also by temporality and its subjective value to the maker. For example, outsourced PCBs often come with a minimum order quantity; thus, although they are technically fully functional, to P10, who only needs two working PCBs, the rest are waste even before arrival. Similarly, intermediate prototypes are useful during design iterations but often end up shelved and eventually discarded. From a systems perspective, this highlights the need for more comprehensive and dynamic approaches to sustainable making—approaches that consider not only the types of waste but also the circulation of waste across the different stages of the making process.

\begin{figure*}[h]
    \centering
    \includegraphics[width=\textwidth]{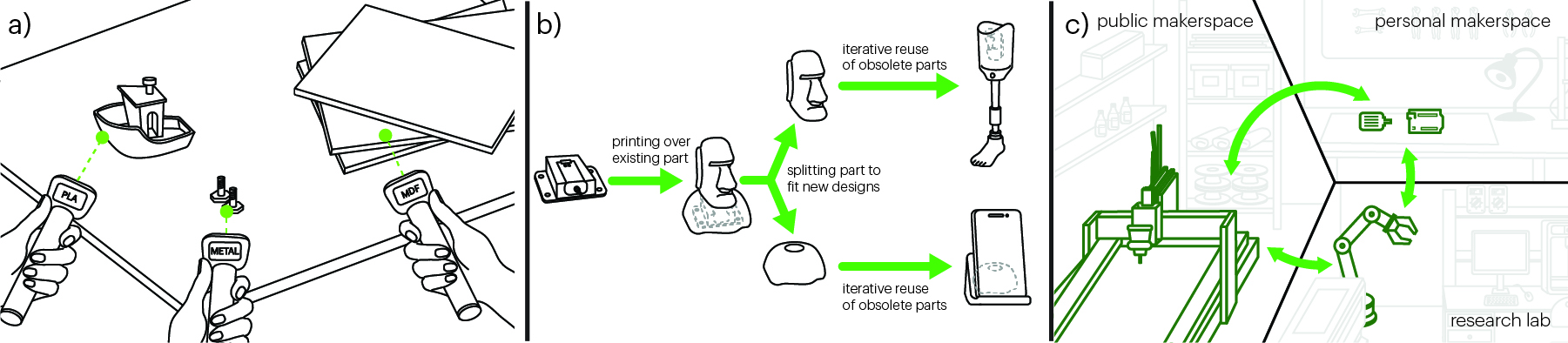}
    \caption{Potential new tools, systems, and infrastructures to support sustainable making. a) Illustration of a handheld material-detection ``torch'', b) illustration of the potential reuse and re-purpose of obsolete 3D printed parts, and c) illustration of a conceptual infrastructure system for resource sharing among makerspaces.}
    \label{fig:discussion}
\end{figure*}

Another key insight pertains to the numerous costs associated with current sustainable practices. As highlighted in Section~\ref{challenges}, sustainability often entails extra burdens, including increased labor, effort, resources, or the need for specialized knowledge. Consequently, it becomes challenging—and sometimes impractical—to advocate for makers to consistently prioritize sustainability in their activities, particularly when they are already constrained by limited time and resources, and when sustainable infrastructure remains underdeveloped.

As members of the technical HCI community and researchers in personal fabrication, we believe that it is critical to provide the maker community with tangible infrastructures that support sustainable making practices. This includes affordable techniques, tools, equipment, and systems that reduce the trade-offs of adopting sustainable practices and, ideally, offer additional incentives for doing so. Such technical innovation should be grounded in the current landscape of making and maker communities, as evidenced by our findings, to ensure that it can effectively address this wicked problem~\cite{pryshlakivsky2013sustainable}. In the rest of this section, we structure our discussion around potential opportunities identified through our findings, which could inform design implications and future directions for scaling sustainable practices. To contextualize these opportunities, we include three illustrations, as shown in Figure \ref{fig:discussion}.

\subsection{Sustainable Disposal of Raw Materials}
As noted in Section~\ref{type_of_waste}, raw material waste is one of the major byproducts of making and is challenging to handle properly. In makerspaces where multiple members share 3D printers, for example, different types of thermoplastics are often discarded indiscriminately in the same plastic bins, as revealed in Section ~\ref{waste_swarf}. Although the intention behind providing plastic-only bins, driven by managers' goodwill, is to facilitate easier recycling, mixing different types of plastics complicates the recycling process. At some sites, such as those of P6 and P12, this issue is addressed by limiting the printing material to PLA-only. However, these rules exemplify the complex trade-off between functionality and sustainability in making practices.

We believe that technical HCI research is well-positioned to investigate such complex, ``wicked'' sustainability challenges. For example, recent work in HCI has developed material-aware solutions for desktop laser cutters, such as a low-cost speckle-sensing add-on that automatically identifies flat sheet materials as they are loaded~\cite{10.1145/3472749.3474733}. 
We speculate that similar techniques, combined with computer vision, could be extended to differentiate 3D printed materials based on their reflective speckle patterns. 
One could envision a handheld ``torch'' with a speckle-sensing array as a low-cost addition to a maker's toolkit, which, by shining light onto a 3D printed object, would allow makers to quickly recognize the raw material of the artifact and efficiently sort 3D printing waste (see example illustration in Figure \ref{fig:discussion}a).

Moreover, similar concepts could be integrated into recycling bins to enable automatic sorting of thermoplastics based on material type. 
This approach could not only improve waste management efficiency but also expand material options at certain sites. Such tools would promote material recycling while offering more flexibility to makers.

Beyond thermoplastics, our findings in Section~\ref{lack_equipment} highlight other common forms of raw material waste that are rarely managed properly due to the lack of procedures and equipment. Sawdust, as one example, was cited by P3 as a problematic byproduct of making—it is difficult to collect, takes up space, lacks centralized processing facilities, and yet is unavoidable. From the manager's point of view (which we also agree with), sawdust differs from common household waste. Disposing of sawdust in a landfill, the current approach, feels wasteful, as it consists primarily of high-quality wood particles and doesn’t require sorting, making it a potential candidate for recycling and reuse with minimal effort. Recent examples in HCI have demonstrated that similar ``pure'' raw material byproducts, such as spent coffee grounds, can be repurposed as eco-friendly 3D printing filament~\cite{10.1145/3563657.3595983}. We believe this exploration of new materials could be further expanded if it were directly integrated into makers' daily activities—for instance, by turning local byproduct waste into new materials for local making.

\subsection{Sustainable Iterative Physical Prototyping}
Our interviews align with much of the HCI literature (e.g., ~\cite{hartmann2006reflective, schon2017reflective}), which describes making as an iterative cycle of ideation and execution, with makers continuously moving between stages and making design decisions throughout the process. Sustainable Interaction Design \cite{SID} emphasizes the importance of placing sustainability at the forefront of each stage. However, our findings indicate that in practice, sustainability is often overlooked. As discussed, rapid physical iteration frequently results in the accumulation of multiple, sometimes unnecessary, intermediate prototypes that are \textit{``too specific to be reused,''} \textit{``piled up in storage rooms,''} and \textit{``eventually thrown away.''}

We argue that these making practices are partly shaped by the affordances of the tools available to makers~\cite{osiurak2016tool,hartson2003cognitive}. The current design-fabrication ecosystem tends to overemphasize the ease of physical creation while offering limited support for activities such as disassembling, updating, or repurposing physical prototypes. Our findings in Section~\ref{optimize_iteration} show that experienced makers often adopt alternative prototyping approaches, such as incremental development locally, rather than constructing the entire prototype at once. This allows for early error detection and prevents the need to recreate the entire prototype. Additionally, breaking a large design into smaller components and iterating only on the uncertain parts proves both time-efficient and effective in reducing the number of physical iterations. Unfortunately, these practices require years of experience and tacit knowledge, which are not well-supported by current design-fabrication tools, making them less accessible to most makers.

This gap highlights new research opportunities in HCI, particularly in designing sustainable fabrication systems that integrate physical design iteration more thoughtfully. 
Several personal fabrication systems have already begun investigating the potential of printing upgraded designs using custom, high-DOF 3D printing hardware (e.g., ~\cite{patchphysical, onthefly, roma}). 
While our work does not aim to exhaustively map these research directions, we identify several promising paths beyond hardware innovation. 
For example, new modeling software could be designed to panelize or split 3D models into standardized shapes—not for ease of printing, but to facilitate reuse in future iterations (Figure \ref{fig:discussion}b). 
This model parsing could be based on the design's semantics~\cite{wang2021hierarchical}, leverage the maker's prior practices, or even take into account the practical realities of the maker's available storage space. 
Modular components would thus be easier to store and reuse across projects (as they are more standardized) while also potentially serving as new infill material for future designs~\cite{10.1145/3411764.3445187}. 
This approach would not only reduce material waste but also lower the energy costs of printing new volumes. Similar concepts could be applied to other common making activities as well, such as new systems that supporting the reuse of electronic components across iterations or between different circuit design projects~\cite{yan2025pcbrenewal, yan2024solderlesspcb, lu2023ecoeda}.

\subsection{Sustainable Knowledge Transfer and Sharing}
While we have discussed ways to better support sustainable practices in making for individual makers, it is clear that viewing sustainable making purely as an individual responsibility is overly narrow. Many of our findings, based on shared makerspace environments, reveal that factors such as physical setups, available resources, types of making, and levels of experience collectively influence makers' decisions to adopt either sustainable or unsustainable practices. This broader context highlights the value of encouraging makers to share their knowledge of sustainable practices more widely, which could serve as an immediate and actionable step toward more sustainable making. Additionally, developing systemic infrastructures to facilitate knowledge and resource sharing across makers and makerspaces could further enhance these efforts.

To build on this, we first emphasize the importance of equipping makers in management roles with a deeper understanding of sustainability. Since makers often follow the frameworks and guidelines set by those in leadership, managers are well-positioned to promote sustainable practices. Our findings indicate that many makerspaces already implement certain sustainability measures, such as setting up recycling stations, labeling components, and encouraging sharing. However, these efforts are often inconsistent and not well-communicated. For example, none of the makers we interviewed recalled receiving specific guidance on sustainability during onboarding, and makers from the same space even had conflicting information about basic practices like waste disposal. This points to an opportunity for managers to introduce sustainability-focused content through structured knowledge transfer processes, including comprehensive onboarding, workshops, or community events.

Beyond managerial roles, sustainability-centered activities can provide valuable opportunities for makers to share knowledge and resources. Prior studies have shown how community engagement can be crucial in driving sustainable making~\cite{10.1145/3232617.3232626, khan2023pragmatics}. Our own research also highlights initiatives such as repair and disassembly sessions, where participants drop off broken machines, and those who repair them get to keep the items. These community-driven engagements not only facilitate knowledge exchange but also promote waste reduction and the efficient use of materials. By leveraging social practices and fostering a collaborative culture, makerspaces can play a key role in promoting sustainability. 

To take sustainability efforts a step further, a more sophisticated approach would be the implementation of a unified information and inventory management system. This system could facilitate the sharing of tools, machinery, equipment, and spare parts across maker communities, reducing waste and maximizing resource utilization (Figure \ref{fig:discussion}c). One specific type of waste we identified involves broken or outdated machines. Often, tools in makerspaces are underutilized, as they may be owned by a single space or person and see limited use. The ``Library of Things'' concept, proposed by Jones et al.~\cite{10.1145/3544548.3581094}, offers a solution—a shared repository of physical items that can be borrowed by the community. While resource sharing is already common within individual makerspaces or institutions, expanding this concept to a broader network presents new opportunities, albeit with added challenges.

Implementing widespread sharing requires more than just inventory management; it demands a cultural shift toward valuing shared resources over personal ownership. Embracing this mindset would allow makers to avoid unnecessary duplicate purchases, which can lead to waste. Additionally, while some makers may lack the skills to repair broken machines, a larger sharing network would bring together individuals with the expertise to maintain and repair equipment, reducing unnecessary waste.

\section{Conclusion}

In this study, we examined the growing sustainability challenges in making as personal fabrication and rapid prototyping continue to expand. Through interviews with grassroots makerspaces and labs, we identified key themes around waste generation, management strategies, motivations for sustainable practices, and challenges faced by makers. While many makers expressed a desire to adopt sustainable practices, practical barriers often prevent them from doing so.

Our research emphasizes the need for infrastructure supporting sustainable making. Our data revealed a scarcity of tools for disassembly, repair, and remaking, highlighting an imbalance between sustainability-focused infrastructure and tools designed solely for making. This gap highlights an urgent need to invest in tools that enable more sustainable practices in makerspaces.

As a final note, the push for local remaking, reusing, and recycling infrastructure may raise additional concerns about whether makers will adopt it and whether it will genuinely promote sustainable practices, i.e., the promotion of new infrastructure can inadvertently introduce unanticipated waste if it is not well-received or effectively utilized. While such concerns are difficult to fully validate, our findings may offer some anecdotal evidence. Recall that P6’s lab progressively adopted a PLA recycling service on campus. Despite material limitations and high maintenance demands, the presence of this service alone positively influenced makers' behavior. When proper waste management services were available, makers became more diligent in sorting waste. Thus, we cautiously advocate for further research, development, and commercialization of sustainable making infrastructure, hoping that a more comprehensive sustainable making ecosystem will ultimately benefit the maker community by fostering more environmentally responsible practices.

\begin{acks}
We sincerely thank all participants for their contributions to this research. We also thank Zining Zhang from UMD for creating the illustrations presented in this paper. An LLM service was used exclusively for proofreading.
\end{acks}

\bibliographystyle{ACM-Reference-Format}
\bibliography{SustainableMakingStudy}
\end{document}